\documentclass[12pt,preprint2]{aastex}
\shorttitle{The present-day MF in NGC\,3603}
\shortauthors{Stolte et al.}

\begin{document}


\title{The secrets of the nearest starburst cluster: \\
II. The present-day mass function in NGC\,3603\footnotemark[1]}

\author{Andrea Stolte\altaffilmark{2}, Wolfgang Brandner}
\affil{Max-Planck-Institute for Astronomy, K\"onigstuhl 17, D-69117 Heidelberg, Germany}
\author{Bernhard Brandl}
\affil{Leiden Observatory, Niels Bohrweg 2, NL-2333 CA Leiden, The Netherlands}
\and
\author{Hans Zinnecker}
\affil{Astrophysikalisches Institut Potsdam, An der Sternwarte 16, D-14482 Potsdam, Germany}
\footnotetext[1]{Based on observations obtained at the ESO VLT on Paranal, Chile,
under programmes 63.I-0015 and 65.I-0135.}
\altaffiltext{2}{\sl send offprint requests to: stolte@astro.ufl.edu}

\begin{abstract}
Based on deep VLT/ISAAC $JHK$ photometry, we have derived 
the present-day mass function of the central starburst cluster 
NGC\,3603\,YC (Young Cluster)\footnotemark[3]
in the giant H{\sc ii} region NGC\,3603. 
The effects of field contamination, 
individual reddening, and a possible binary contribution are investigated.
The MF slopes resulting from the different methods are compared, and lead 
to a surprisingly consistent cluster MF with a slope of $\Gamma = -0.9 \pm 0.15$.
Analyzing different radial annuli around the cluster core, no significant change 
in the slope of the MF is observed.
However, mass segregation in the cluster is evidenced by the increasing 
depletion of the high-mass tail of the stellar mass distribution with increasing
radius. We discuss the indications of mass segregation with respect to the 
changes observed in the binned and cumulative stellar mass functions, 
and argue that the cumulative function as well as the fraction of high-
to low-mass stars provide better indicators for mass segregation than 
the MF slope alone. Finally, the observed mass function and starburst 
morphology of NGC\,3603\,YC is discussed in the context of massive local 
star-forming regions such as the Galactic Center Arches cluster, 
R\,136/30 Dor in the LMC, and the Orion Trapezium cluster,
all providing resolved templates for extragalactic star formation.
Despite the similarity in the observed MF slopes, dynamical considerations
suggest that the starburst clusters do not form gravitationally bound systems
over a Hubble time. Both the environment (gravitational potential of the Milky Way) 
and the concentration of stars in the cluster core determine the dynamical stability 
of a dense
star cluster, such that the long-term evolution of a starburst is not exclusively
determined by the stellar evolution of its members, as frequently assumed for 
globular cluster systems.
\end{abstract}
\footnotetext[3]{As defined in Stolte et al.~2004 (Paper I), we refer to the 
central cluster in NGC\,3603 as ``NGC\,3603\,YC'' to avoid confusion with 
the extended H{\sc II} region.}
%


\section{Introduction}

High-mass star formation is still a puzzle to theorists and 
observers alike. Star-formation theory suggests that high-mass
stars ($M > 20\,M_\odot$) cannot form in the standard 
fragmentation and subsequent accretion scenario without 
serious modifications to the physical processes involved
\citep{Stahler2000}.
Competitive accretion \citep{Bonnell2001}, merging of protostellar
clumps or young stars (Bonnell et al.~1998), and enhanced accretion
in dependence on the star-forming environment \citep{Behrend2001}
are all scenarios suggested for the formation of O-type stars.
All these processes share the requirement of exceptionally high 
gas and/or stellar densities to create high-mass stars.

The necessity of a high-density environment where altered
physical processes shape the resultant stellar mass, and predominantly
create high-mass stars at the cost of low- or intermediate mass objects,
predicts primordial mass segregation in the densest star-forming loci, 
which might be evidenced in a flattened initial mass function (IMF)
with a strong bias to high-mass stars in comparison to more 
moderate sites of star formation. In contrast to this, the IMF
is found to be surprisingly invariant in a diversity of Milky Way 
and Magellanic Cloud star forming regions \citep[b]{Massey1995a}.

The densest environments available to study the stellar mass spectrum
are starburst clusters. In the Milky Way, only very few compact, dense
starburst clusters are known, with the youngest and most condensed systems being 
the Arches cluster close to the Galactic Center (GC) and the central starburst
cluster in NGC\,3603. These clusters are particularly important 
for our understanding of extragalactic star-forming regions, where 
the starburst mode is the predominantly observed mode of star formation 
due to the intrinsic brightness of compact, massive young clusters. 
Star-forming regions in other galaxies 
are, however, in most cases too distant to be resolved into individual stars.
A detailed study of starburst clusters in the Milky Way 
provides a resolved template for extragalactic star-forming regions.

With the aim to observe possible deviations or to confirm the 
universality of the initial mass function in starburst environments, 
we derive the stellar mass distribution 
in NGC\,3603\,YC (Young Cluster). We analyze possible changes in 
the MF slope, in particular with respect to the distance from the 
cluster core, where the highest densities are found.
A possible deviation from a standard MF in starburst environments
would have severe consequences on our understanding of star formation 
in the nuclei and tidal interaction zones of distant galaxies,
such as e.g. the Antennae galaxies, and star formation in the early 
universe when the star formation efficiency was more intense. 
This latter aspect refers not only to the distant universe, but also 
to the formation epoch of globular clusters in the Milky Way.

As the NGC\,3603\,YC starburst cluster is extreme among today's 
Milky Way massive star-forming regions, and with a stellar cluster mass of 
$\sim 10^4\,M_\odot$ residing in a cloud of total gas mass $4\cdot 10^5 M_\odot$ (Grabelsky
et al.~1988) at the low-mass end of extragalactic starbursts, 
several attempts have been made to understand the stellar mass function.
In all these studies, the resolution in the core was (and still is)
the limiting factor, causing large uncertainties in the resultant MF shape.
From optical photometry, Moffat et al.~(1994) obtained a MF slope of
$\Gamma = -1.4 \pm 0.6$ for $30 < M < 60\,M_\odot$, i.e., the 77
brightest stars. Hofmann et al.~(1995) derived $\Gamma = -1.6$ in the 
mass range $15 < M < 50\,M_\odot$ from only 28 central stars 
resolved with speckle interferometry. While these slopes are 
consistent with a Salpeter MF, only the massive stellar population
is traced. The limitations in resolution additionally constrain
these results mostly to the outer cluster regions (except for the 
speckle data, where statistics are very poor). 
The earliest attempt towards a complete central MF was performed by
Eisenhauer et al.~(1998), who derived a MF slope of
$\Gamma = -0.73$ for $1 < M < 85\,M_\odot$ in the innermost
$13^{''}$ of the cluster center from ESO/ADONIS high-resolution
adaptive optics photometry. In this study,
crowding effects were taken into account via 
spatially varying incompleteness ratios, and the presumed age spread 
of $0.3 - 1$ Myr on the PMS was considered
by calculating a median age of the young, intermediate mass 
population of 0.5 Myr from individual positions of PMS stars in the 
color-magnitude diagram.
Although these authors give strong arguments for 
a young main-sequence cluster age of only 1-2 Myr, they use a 
3 Myr Geneva isochrone to derive the MF at the high-mass end.
Recently, Sagar et al.~(2001) confirm this slope with 
$\Gamma = -0.84$, however from ill-resolved $UBVRI$ data, such that, 
again, the innermost cluster region could not be traced. 
Despite these uncertainties in previous studies, high-resolution $VRI$ HST/PC data 
(FWHM $< 0\farcs1$) yield a similarly flat slope of $\Gamma = -0.9 \pm 0.1$ 
in the mass range $1 < M < 100\,M_{\odot ,initial}$ in the inner $20^{''}$
of the cluster center using a composite isochrone for the MS and PMS 
population with a single age of 1 Myr \citep{Sung2004}.

We will use the cluster characteristics determined in Stolte et al.~2004
(hereafter Paper I, where we give a detailed introduction on NGC\,3603;
see also the introduction in Sung \& Bessel 2004 for a review on the cluster),
in particular a uniform cluster age of 1 Myr and a distance of 6 kpc (Paper I), 
to derive mass functions and possible radial variations in the MF from 
VLT/ISAAC near-infrared data (see Paper I, and Brandl et al.~1999) 
obtained under excellent seeing conditions (FWHM $< 0\farcs4$). 
This data set contains the deepest currently available photometry of 
NGC\,3603\,YC, allowing one to observe cluster members down to masses of 
$\sim 0.1\,M_\odot$, well into the subsolar regime. 

In Sec.~\ref{obssec}, we give a brief summary of the observations
presented in Paper I.
The use of high-resolution near-infrared data allows us to derive the 
MF slope down to $0.4\,M_\odot$ in the crowding-limited cluster field,
while at the same time minimizing effects from variable extinction.
We recall the critical cluster properties used in the MF derivation 
in Sec.~\ref{clusprop} from Paper I.
The mass functions are derived in Sec.~\ref{mf3603sec}, and the resultant
slopes are discussed in Sec.~\ref{mfdiscsec}. Radial variations in 
the MF slope are analyzed in Sec.~\ref{mfradsec}, and the effects 
of mass segregation are interpreted in Sec.~\ref{massseg}.
Because of the impact of mass segregation on our understanding of 
the star formation process, we present a detailed comparison between 
different star-forming environments in Sec.~\ref{comparesec}.
Finally, we summarise our main results briefly in Sec.~\ref{sumsec}.

\section{Observations}
\label{obssec}

A detailed analysis of the photometry of NGC3603\,YC is presented in 
Stolte et al.~(2004, Paper I), where color-magnitude and color-color 
distributions of the cluster center were derived and discussed.
For the sake of completeness, a brief summary of the observations
contributing to the derivation of the mass function is given here. 
Deep VLT/ISAAC observations were taken in the $J_s$, $H$, and $K_s$ bands.
The $J_s$ and $K_s$ observations resulted in the best resolution 
due to stable observing conditions, while the seeing degraded and 
varied during the $H$-band observations. The $J_s$ and $K_s$ data
display the highest spatial resolution on the final images, 
thus yielding the best photometric performance, and will therefore form 
the basis for the mass function derivation. In order to avoid
artefacts and nebulous peaks, it was nevertheless required that
each object in the final photometric catalogue be detected in all
three filters, including $H$. 

With the goal to resolve the cluster center, two sets of images were
created. As crowding (as opposed to photometric depth) is the most limiting 
factor in the cluster center, the lowest seeing images were combined to 
$J_s$ and $K_s$ integrations of 9 and 11 min, respectively, 
covering a $2^{'} \times 2^{'}$ field of view centered on the cluster. 
The final resolution on the combined images is $0.\!^{''}38$ in $J_s$
and $0.\!^{''}35$ in $K_s$.
In addition, the total set of 35 images in $J_s$ and 39 images in $K_s$
yielded combined frames of 35 and 39 min exposure time and a field of view
of $3.\!'4 \times 3.\!'4$, with final resolutions of $0.\!^{''}40$ and
$0.\!^{''}37$, respectively. This larger field is used to estimate the field
contamination in the cluster sample from the edge of the field of view,
where crowding does not affect the photometry. The brightest stars in the field
and in particular the bright core of NGC\,3603\,YC inside a radius of $R=7^{''}$
suffer from severe saturation, such that the core itself had to 
be excluded from the analysis. As the most massive stars reside inside
the core, this implies that the upper end of the mass function cannot 
be derived from these data. As no NIR data set with comparable resolution
covering the larger cluster area is available at the present time,
we cannot easily fill in the high-mass end.
Analysis of the upper end of the mass function in the cluster core
as determined from high-resolution adaptive optics
or HST/WFPC2 data is discussed elsewhere \citep{Eisenhauer1998,Sung2004}, 
and the results from these studies will be included in the discussion 
of the mass function.

The details of the observations can be found in Paper I, Sec. 2, Tabs. 1 and 2.

\section{Cluster properties of NGC3603\,YC}
\label{clusprop}

In Sec.~\ref{mf3603sec}, various methods will be used to derive
the mass function in NGC\,3603\,YC. Before the methods suggested
from the complex physical properties observed in the cluster
center are described (cf. Paper I),
the main physical parameters such as the distance, the age
and the metallicity of NGC\,3603\,YC are recalled from Paper I. 

\subsection{Distance, extinction and age derived from the PMS transition region}

The distance, extinction and age of the starburst cluster were determined
from the ISAAC data via isochrone fitting to the pre-main sequence/main sequence 
(PMS/MS) transition region in Paper I. Palla \& Stahler (1999) pre-main sequence
isochrones of ages 0.3-3 Myr were used to perform the fit. The selection
of isochrones is discussed in Paper I. As these parameters are crucial 
for the mass function derivation, we summarise the results below.

\subsubsection{Distance to NGC 3603}

Kinematic distance estimates to NGC\,3603\,YC range from 6 to 10 kpc,
while early photometric studies yielded $\sim 7 \pm 0.5$ kpc
\cite{Moffat1983,Melnick1989}. The purely empirical distance of 7 kpc
derived by Moffat (1983) from optical photometry and spectral 
classification of massive cluster members was essentially used in 
all subsequent studies. Although independent of stellar evolution 
models, these early photometric distance determinations were restricted
to the main sequence population.

We obtain a distance modulus of $DM = 13.9 \pm 0.1$ mag from 
isochrone fitting of the PMS/MS transition 
region. The corresponding distance, $d = 6 \pm 0.3$ kpc, 
is slightly lower than previously derived values,
but consistent with the kinematic distance estimate of $6.1 \pm 0.6$ kpc 
from De Pree et al.~(1999), and the photometric distance of 
$6.3\pm 0.6$ kpc derived by Pandey et al.~(2000).

\subsubsection{Foreground extinction to NGC 3603}

The foreground extinction as derived from spectra of high-mass 
stars close to the cluster center is $A_V=4.5 \pm 0.3$ mag \citep{Moffat1983},
and $A_V = 4.6\pm0.6$ from optical photometry \citep{Melnick1989}.
Melnick et al.~1989 observe a radial extinction variation, confirmed
by Pandey et al.~2000, which suggests an extinction of 4 mag 
($E_{B-V}\sim1.3$, albeit uncertain due to large scatter)
when extrapolated to the cluster center\footnote{A radial variation 
may be an oversimplification, especially because IRAS maps indicate 
a North-South orientation of dense material around the cluster 
\citep{Nuern2002b}. The insensitivity of NIR data prohibits 
a more accurate determination of the spatial extinction variation.}. 

A foreground extinction of $A_V = 4.5\pm0.6$ mag is derived from PMS isochrone 
fitting towards the pre-main sequence population in Paper I, consistent with 
earlier studies. 
The large uncertainty mirrors the low sensitivity of infrared 
observations to granular material along the line of sight.
This choice also fits the lower PMS stars towards fainter magnitudes, 
beyond the distinct, horizontal transition region. 
A slightly lower extinction of $A_V = 4$ mag is derived for main sequence stars,
pre-dominantly found close to the cluster core,
slightly lower than extinction estimates from spectroscopic studies, where values 
around $A_V=4.5$ mag were suggested \citep{Moffat1983}, but consistent with the 
core extinction expected from the radial trend observed by Melnick et al.~1989. 
In this estimate the assumption was made 
that the resolved secondary sequence is comprised of a physically distinct
population (due to binarity or other physical processes), 
and therefore excluded from the isochrone fit. Objects on the 
secondary sequence display redder colors, such that the extinction estimate 
increases when the whole population is considered, in agreement with earlier
results. \footnote{Note, however, that increasing color terms for redder 
PMS stars could not be taken account due to the limited number of standard
stars. Thus, we cannot entirely rule out an instrumental color effect 
between MS and PMS, or uncertainties in the extinction law or PMS isochrones, 
to cause the derived deviation in $A_V$. See Paper I for a detailed discussion 
of the uncertainties.}
As stars will be dereddened individually in the MF derivation, which implies 
shifting stars along the reddening path to $A_V=0$ mag, the difference
in extinction between PMS and MS is meaningless for the mass calculation.

\subsubsection{Age of the starburst cluster}

The age of NGC\,3603\,YC has evolved in the literature during the past
20 years. Early studies suggested an age of 2-3 Myr from the fact that 
presumably evolved stars with strong winds and Wolf-Rayet characteristics 
reside in the cluster center \citep{Moffat1983}. 
In the past decade, however, it became increasingly clear
that extremely massive stars of type O2 show strong Wolf-Rayet characteristics
while still on the main sequence (see, e.g., Walborn et al.~2002). 
A main sequence nature is evidenced in the spectra of the most massive stars 
in NGC\,3603\,YC by strong hydrogen emission \citep{Drissen1995,Drissen1999}. 
Moffat et al.~(2002) argue that the WR stars in NGC\,3603\,YC
are ``probably main sequence stars of extremely high luminosity
with strong wind-produced emission lines'' from 
the correlation observed between X-ray and radio flux,
which is uncorrelated for normal Galactic (single) WRs.
Massey \& Hunter (1998) compare the spectra
of three WRs in NGC\,3603\,YC to similar stars observed
in R\,136/30 Dor, and conclude that these are probably ``super O stars''. 
From the short main sequence lifetime of these massive beasts, the estimated cluster age 
has decreased in the literature to about 1 Myr \citep{Drissen1999,Moffat2002},
which is consistent with age estimates indicated by PMS isochrones
\citep{Brandl1999,Eisenhauer1998}.

In Paper I, we obtain an age of 1 Myr as the best fit to the PMS/MS transition 
region. The significant width of the transition region caused the suggestion of 
an age spread between 0.3 and 1 Myr in the starburst cluster in these previous 
studies. As discussed in 
detail in Paper I, there are two straight-forward interpretations of the width 
of the transition region. In addition to an age spread, the PMS transition 
can also be interpreted as comprised of a mixture of single and binary stars, 
with the well-populated upper envelope indicating binaries with mass ratios close 
to unity. This interpretation is supported 
by a secondary sequence observed close to the main sequence in the ISAAC data.
From $\chi^2$ distance minimisation, the best fit to the PMS population 
is obtained for a single cluster age of 1 Myr, including the assumption that
stars on the secondary sequence are binaries. An age distribution can, however,
not be entirely excluded prior to a detailed spectral analysis of objects in 
this regime. For the mass function derivation, we will 
therefore assume a single cluster age of 1 Myr, while treating the secondary 
sequence in different ways to minimise a potential bias introduced by the 
interpretation of the secondary sequence, e.g.~as binaries.

As stars on the MS do not evolve severely, slight deviations
in the age determination are not critical for the derivation
of the main sequence MF ($M > 4\,M_\odot$ according to a 1 Myr 
Palla \& Stahler (1999) isochrone). 
At these young ages, stellar evolution is almost negligible except
for the highest mass objects, which are not included here,
such that the mass function derived from a 1 Myr MS isochrone should not 
deviate significantly from earlier MFs using a 
3 Myr model. The main sequence portion of the MFs derived in the 
following is therefore comparable to the results of previous studies,
while the pre-main sequence MF depends strongly on the assumed 
age and the treatment of stars in the transition region.

The isochrones used to determine masses of individual stars from 
the color-magnitude plane will be shown along with the PMS and MS selection 
of stars in Sec.~\ref{mf3603sec}, Fig.~\ref{pmsmssel} (see also Paper I).

\subsubsection{Metallicity of NGC 3603}

No thorough analysis of NGC\,3603 with respect to metallicity 
is available in the literature. The spectral analysis of the WR-like 
massive component in the cluster core suggests a metallicity close 
or equal to solar \citep{Schmutz1999}. 
NGC\,3603 is located at a Galactoccentric
distance of 8 kpc, comparable to the Sun. The radial gradient in 
metallicity observed in the Milky Way supports that the metallicity 
in NGC\,3603 is comparable to the solar value, which was adopted.

\subsection{Uncertainties in the mass function derivation}

The color and brightness of stars at the youngest evolutionary stages are 
influenced by a diversity of processes. Variable extinction can be 
caused by circumstellar envelopes or a variation in the density
of remnant molecular material along the line of sight.
Near-infrared excess emission from circumstellar disks alters the
colors and increases the apparent reddening of stars, thereby causing
an overestimate in the stellar mass of individually dereddened objects
with intrinsic NIR excesses.
Binarity/multiplicity increases the brightness, again resulting in 
an overestimate of the stellar mass. 
The large scatter observed in particular at fainter magnitudes
in the PMS/MS transition and PMS region of the CMD suggests that 
these evolutionary stages are especially affected, although the increasing
photometric uncertainty toward fainter magnitudes prohibits a quantitative 
analysis of increasing NIR excesses in PMS vs. MS stars. These uncertainties
for masses below $4\,M_\odot$ have to be kept in mind during the derivation
and interpretation of mass functions.

While variable extinction can 
be taken into account by individually dereddening stars to the 
theoretical isochrone, NIR excess and binarity are difficult to assess.
As the majority of stars in NGC\,3603\,YC are found close to the 
fitted isochrone, severely reddened sources, either excess cluster members 
or background objects, can be excluded from 
the MF derivation without significant statistical loss.
Multiplicity can usually not be taken into account in MF derivations.
With the interpretation of the secondary sequence in NGC\,3603\,YC as 
a sequence of binaries with mass ratios close to unity, we can, however, estimate 
the contribution of (equal-mass) binaries to the MF and correct for this effect.

\section{The Mass Function in the central NGC\,3603 cluster}
\label{mf3603sec}

\subsection{General remarks on the mass function derivation in NGC\,3603\,YC}

As a consequence of the clear separation of a pre-main sequence and 
a main sequence population observed in the central CMD of NGC\,3603\,YC, 
and the existence of a visible secondary sequence interpreted as binary 
candidates (see Paper I), the mass function derivation has
to incorporate a number of physical assumptions.
While generally one favourite set of stellar evolution models is chosen to 
derive the mass function via isochrone fitting, in the case of NGC\,3603\,YC
two different sets of evolutionary models are required to represent
the MS and the PMS population at a given age. As the 1 Myr pre-main sequence
isochrone from the Palla \& Stahler set of models fits both the transition
of the PMS to the MS as well as the lower PMS distribution best (cf. Paper I), 
this isochrone is chosen for the transformation of luminosities into masses on the PMS. 
For the evidence given above and in Paper I, the MS population 
is also consistent with a single star-forming burst 1 Myr ago. Along with 
the binary interpretation, this suggests a single age for the entire 
cluster population, without the need to employ an age spread.
The mass-luminosity relation for main sequence stars is deduced from
a 1 Myr solar metallicity isochrone from the Geneva basic grid of stellar 
evolution models \citep{Lejeune2001}. Effects such as enhanced 
mass loss are not taken into account, as the ISAAC data are saturated
for stars with masses above $20\,M_\odot$, below which mass loss is 
not significant at these young ages.
In addition, the case of a 2 and 3 Myr MS isochrone, used in 
prior MF derivations and allowing for some uncertainty in the derived age,
will also be tested to deduce potential effects of stellar evolution 
on the resulting MF slope.

The identification of a secondary sequence poses an additional challenge to a
realistic MF derivation. As no prior data set of the starburst cluster 
resolved this feature clearly,
the MF will be shown with and without binary correction to allow comparison
with prior MF derivations. 

As a consequence of the variable extinction observed in the NGC\,3603 region, 
individual dereddening was applied to determine stellar masses. 
As the NIR data alone do not allow us to derive the extinction law 
toward NGC\,3603, the extinction law was adopted from Rieke \& Lebofsky (1985).
Each star has been shifted along the reddening vector in the 
$J_s,\ J_s-K_s$ plane onto the PMS or MS isochrone.
Although this procedure is more accurate in the sense that individual
reddening is taken into account, it implicitly assumes that all scatter 
observed in the CMD is caused by reddening variation. This is clearly not 
the case. Besides the photometric scatter inherent in the data,
other processes such as binarity, stellar rotation, and intrinsic IR 
excesses contribute to the scatter. Most of the additional processes influencing a 
very young stellar population are difficult to quantify, and cannot be taken 
into account in the following analysis. 
While photometric uncertainties are expected to introduce random 
scatter, such that their combined effect should be negligible, 
individual dereddening allows one to estimate the maximum influence of 
extinction variations on the MF slope.
The intense crowding in the cluster center decreases the detectability
of lower-mass stars and can cause a systematic uncertainty in the MF slope,
which would mimic a bias to high-mass stars.
For each cluster area/annulus the average
incompleteness is therefore estimated as a function of the radial distance
from the cluster center and hence stellar density.

While a detailed analysis of the uncertainty in the MF slope 
imposed by the choice of stellar evolutionary models is beyond 
the scope of this paper, a thorough discussion on this subject
can be found in Andersen (2004).

The procedures used to derive the stellar mass distribution in 
NGC\,3603\,YC are outlined in detail in the next sections.
%

\subsection{Incompleteness correction}

Artificial star tests were performed on each science frame to 
determine the recovery fraction per magnitude. Between 200 and 250
stars were inserted in each artificial frame to avoid enhancing crowding
effects. 45 artificial frames were created from the smaller 
$2^{'} \times 2^{'}$ ``cluster'' images and 30 artificial frames from 
the larger $3.\!^{'}4 \times 3.\!{'}4$ ``field'' images.
The inserted number of stars corresponds to less than 1\% of the total
number of stars detected on each image. Magnitudes and positions
were assigned randomly using the {\sl addstar} task in IRAF. 
As the cluster itself covered 
only a small area on the high-resolution frames, additional simulations
were performed on the innermost $60^{''} \times 60^{''}$, with 
100 to 150 artificial stars or $\le 2\%$ of the stellar population found
in the cluster center. Source detection and PSF fitting was conducted with 
the same parameters as for the stellar photometry. The procedure was
repeated to yield the recovery fractions for $\sim 6000$ stars in the field
frames, and $\sim 11000$ stars in total on the cluster frames, with
6000 stars distributed randomly over the entire image, and 5000 stars
confined to the critical central region. The resultant recovery fractions 
were applied in two steps during the analysis: 1) during field star correction
and 2) during incompleteness correction of the mass function.

\subsection{Field star correction}

%
The field star fraction was derived from an area to the North and East
of the cluster center.
This region is particularly well suited to estimate field stars
in the low-extinction cluster center, as it is neither influenced 
by strong H{\sc ii} region emission nor shows evidence of variable 
or enhanced extinction. Field stars were statistically subtracted in each $0.5 \times 0.5$
mag color-magnitude bin of the CMD. Prior to statistical field star subtraction,
the number of field stars had to be corrected for the varying stellar 
density (crowding) with radial distance from the cluster center.
For this purpose, the field and cluster recovery fractions, $f_{field}$ and $f_{cluster}$,
were fitted as a function of the magnitude as a third order polynomial
as displayed in Fig.~\ref{n3603incfit}.
The number of field stars is corrected for field incompleteness as 
$n_{field,corr} = n_{field}/f_{field}$, i.e. to a field completeness
of 100\%. This number was then reduced according to the expected 
recovery fraction of stars in the radial annulus studied, yielding a combined
correction of $n_{field,final} = f_{cluster}\cdot n_{field,corr} = f_{cluster}/f_{field}\cdot n_{field}$, with $f_{field,final} = f_{cluster}/f_{field}$ corresponds to the final density/crowding 
correction factor applied to adjust the field star number counts to the cluster center.
In addition, the measured number of field stars on the larger field area 
was scaled down to the cluster or annulus area considered.

%
The incompleteness simulations yielded independent correction factors for 
each filter. In principle, the combined probability to find a field star
in each color-magnitude bin in the $J_s,\ J_s-K_s$ plane is then given by the product
of the individual probabilities, $f_{JK} = f_J \cdot f_K$. In the color and
magnitude range covered by the cluster center stars, however, the incompleteness
is entirely determined by the $J$-band limit. Thus, only $J$-band incompleteness
is taken into account in the mass function derivation. The correction factor
$f_J$ for each magnitude is obtained from the polynomial fit. The measured
number of field stars is corrected by the factor derived for each annulus
studied, and the adjusted number of field stars is then statistically removed
from the cluster CMD.

As the extent of the cluster is debated, with estimates ranging up to a radius
of $2.\!^{'}5$ (N\"urnberger \& Petr-Gotzens 2002), 
the area covered by the ISAAC data may overestimate 
the field star contamination. As the low-mass halo surrounding NGC\,3603\,YC
is mainly composed of faint, low-mass stars, the bright central
cluster population should not be severely affected\footnote{Ideally, a separated
field should be used to estimate the background population. One off-field was 
observed at a distance of $\sim 30^{'}$
from the cluster center, but unfortunately turned out to contain a
different stellar population dominated by older giant stars. The strong spatial
variations in the stellar population along the Carina arm complicate the 
determination of the field star contribution in any particular region.}.

Fig.~\ref{cmd_n3603} shows the effects of field subtraction on the central
cluster population, and also indicates the 50\% completeness limit for 
this radial selection, down to which mass function slopes will be derived.

%
In the following sections, the different approaches used to derive the 
mass function in NGC\,3603\,YC will be described in detail. 
Results for all MF derivations are summarised in Tab.~\ref{n3603mftab}.
The combined PMS/MS mass functions of this selection are fitted
in Fig.~\ref{mf_all}.
The apparent MS population fainter than $J_s = 15.5$ mag, 
corresponding to the 1 Myr MS turn-on at $\sim 4\,M_\odot$, 
is likely comprised of MS stars in the NGC\,3603 region. This population
has to be far older than the cluster age, and cannot originate in the 
cluster. The lower MS beyond the cluster MS turn-on 
is therefore excluded from the fit. Thus, the combined low-mass MF 
below the MS turn-on at $M=4\,M_\odot$ is solely comprised of PMS stars.

\subsection{Simple star counts}

In the simplest model, the luminosity of a star reflects its
evolutionary stage and mass. This ignores all processes changing
the luminosity and thus magnitude of a star individually and independently
of its evolutionary stage. The isochrone can then be applied directly
to each given stellar magnitude to transform the brightness distribution 
into a mass function. As this is the most
frequently used method to derive mass functions, we include it here as 
the simplest approach. Field data are not always available, and field
contribution can indeed be negligible in the dense cluster regions,
such that field star contamination has rarely been taken into account in MF 
derivations of the NGC\,3603\,YC (except for the study of the 
cluster center by Eisenhauer et al.~1998).
In order to show the effects of field contamination alone,
MFs were calculated from simple star counts both
with and without field subtraction. \\

\subsection{Individual dereddening}

The stellar population in NGC\,3603 shows individually varying extinction, 
caused either by variations in the distribution of granular material in the 
foreground, or by individual reddening from remnant circumstellar
envelopes. The maximum possible effect of individual reddening on the 
MF slope can be probed by assuming that the photometric scatter is 
caused by varying extinction alone. Individual reddening is then taken 
into account via dereddening each star along the reddening vector 
(Rieke \& Lebofsky 1985)
in the $J_s$,$J_s-K_s$ plane until an intersection with the corresponding
isochrone is reached, and the mass of the star is determined from 
the intersection point. The extinction applied to the isochrone itself
is not relevant in this case, as only the length of the dereddening
path (i.e., the relative shift in the CMD) depends on the location of 
the isochrone, but not the desired intersection point. For the 
MF derivations below, isochrones matching the average foreground extinction
of the MS and PMS populations were used, and stars blueward and redward
of the isochrone have been shifted accordingly. The blue and red limits
in the distance from the isochrone was determined by eye from the stellar
distribution in the $J_s, J_s-K_s$ plane. As the main sequence is not
affected by field star contamination, all stars around the MS are 
included in the MF, with a color selection of -0.25 mag blueward and
+0.35 mag redward of the MS isochrone. The bulk of the PMS population,
where reddening is more severe, can be included selecting colors -0.25 mag 
blueward and +0.5 mag redward of the PMS isochrone. This rejects the 
reddest objects ($A_V > 7.5$ mag), which might be affected by 
infrared excesses and background contamination.
The selection of stars contributing to the mass function is shown in 
Fig.~\ref{pmsmssel}.

%
Note that most of the stars are found in a very narrow range around 
the PMS, in an envelope of only -0.1 to +0.15 mag. This narrower selection does
not alter the slope of the PMS MF alone, but in comparison with the MS
MF too few stars are found in the transition region, such that MS and PMS
do not agree at the transition mass. Therefore, the wider PMS selection was 
used, resulting in very good agreement of the MS and PMS MF bins
at the transition mass. \\

In all following MF derivations, field subtraction and individual reddening
are taken into account.

\subsection{Treatment of the PMS/MS transtion region}

Stars in the transition region between PMS and MS 
were allocated PMS transition masses, which are in the narrow 
range between 2.5 and $4\,M_\odot$, depending on their distance 
to the isochrone. No individual dereddening was performed in this 
case due to the large scatter in the transition region.
There is a slight ambiguity in the 
Palla \& Stahler isochrone in the sense that stars with 3.5 to $4\,M_\odot$
cannot be distinguished, but this is not relevant for the resultant MF
as these stars all fall into the same mass bin. 
Only masses above the hydrogen 
burning turn-on of $4\,M_\odot$ enter the MS MF.
Although this facilitates the analysis as a clear cut is introduced
between MS and PMS contributions, the transition region is clearly
physically more complex than envisioned in this simple treatment.
In particular, stars contributing to the MS may be scattered away 
from the narrow range beyond the PMS transition point in the CMD, rendering 
the respective MS and PMS contributions close to the $4\,M_\odot$ PMS 
limit very uncertain. The transition mass
will be marked in all derived MFs, such that the combination 
of MS and PMS contributions is clear, and in particular the last PMS and 
first MS bin should be observed with care.

\subsection{Treatment of the secondary sequence}

Evidence for a binary nature of a secondary sequence observed
in the ISAAC CMD of NGC\,3603\,YC is given in Paper I.
However, this interpretation is as yet unproven by spectroscopy.
Three different approaches are therefore chosen in the treatment
of the secondary sequence. 1) The secondary sequence is 
completely rejected from the mass function derivation. As the relative
fraction of stars on the secondary sequence is not observed to vary 
with stellar brightness within the statistical uncertainties, removing
stars on the secondary sequence corresponds to statistically subtracting
$\sim 30$\% of the population above $2.5\,M_\odot$, below which the vertical
turn of the PMS does not allow one to distinguish a secondary sequence. 
While this may result
in too low star counts in high-mass bins with respect to the low-mass MF,
the statistical properties and thus the slope in the high-mass MF 
should not be altered. 2) The distinct interpretation
is ignored and the entire population is used for the MF derivation,
in agreement with earlier studies. This MF is also directly comparable
with MF derivations that do not have the photometric accuracy
or resolution to attempt a binary correction, such as the early 
study by Salpeter (1955). 3) The binary interpretation
is used to correct for the effect of binaries with mass ratios close 
to unity on the resultant MF slope. \\

\subsubsection{Binary rejection} 

The selection of stars on the secondary sequence was shown and 
carefully discussed in Paper I, see esp.~Fig.~7.
The band of stars above the transition region and redward 
of the main sequence was rejected in the first estimate of the MF,
assuming that some currently unknown physical process causes the 
offset in the CMD, and thus masses may not be accurately determined.
If stars on this sequence are not subject to a mass bias 
from a different formation process, the resultant ``single'' 
mass function should represent the underlying statistically-unbiased
stellar population of the star cluster. 
The vertical distribution of lower PMS stars does not 
allow such a selection. Even if the same photometric offset is
included in the case of PMS stars, as is to be expected from 
the secondary sequence extending all the way into the transition region,
such an offset is invisible in the vertical PMS feature.
As a consequence, the corresponding PMS stars cannot be rejected,
and only the MS and the transition region (down to $2.5\,M_\odot$) are
affected by binary rejection.

\subsubsection{Binary correction - MS and transition region}

As a first step toward a binary corrected MF, 
the visible binary candidate sequence above
the main sequence and the transition population was 
adjusted. The $J_s$-band magnitude of the selected stars
was increased by 0.75 mag as expected for equal-mass binaries, 
and an additional companion star with equal color and magnitude
was included in the number counts. 
This procedure makes the simplifying assumption
that all binaries contributing to the brighter secondary 
sequence are close to equal-mass systems as suggested 
from the observed magnitude offset,
in accordance with the observational restriction that 
for arbitrary mass ratios no prediction can be made. 
Thus, the MFs derived below 
are still system MFs uncorrected for companions with masses
significantly lower than the primary mass. 
The adjustment is applied after field subtraction, before
individual dereddening and MF calculation. This way, the
bias from individual dereddening of equal-mass binaries 
due to the larger brightness and thus higher mass estimate is avoided.
The resultant pairs of ``single'' stars are treated in exactly the 
same way as the rest of the population.

\subsubsection{Binary correction - statistical PMS correction}

While a secondary sequence is observed above the MS and the 
PMS transition region, the problem of the merging of binaries into
the lower PMS population remains. The fraction of stars observed
on the secondary sequence is approximately 30\% of the total 
number of stars both parallel to the MS and in the PMS/MS transition 
region ($J \le 15.5$ mag). When 
intervals of 1 mag width are studied, no trend is observed
(Tab.~\ref{binfractab}) toward fainter stars. 
Thus, to correct for the equal-mass binary contribution 
on the PMS a constant binary fraction of 30\% has been assumed. 
If no binary formation mechanism enhances the number of high-mass 
binaries or decreases the number of low-mass binaries with mass ratios
close to unity, 
a significant fraction of binaries should be hidden in the PMS distribution.
In order to correct for the bias imposed
on the derived MF, 30\% of the PMS stars were selected 
randomly, shifted down by +0.75 mag and an additional 
companion was added, as in the case of the MS binary sequence.
If the equal-mass binary fraction is the same for each mass
bin, as suggested from the MS and transition region,
this procedure should statistically correct for the overestimate of 
individual masses due to brighter magnitudes of potential binaries.
%

\section{Discussion of the resultant mass functions}
\label{mfdiscsec}

Mass functions were created from the above procedures with a 
histogram bin width of $\log(M/M_\odot) = 0.2$, starting at $0.1\,M_\odot$.
A linear least-squares fit was applied to the incompleteness corrected 
MF from the 50\% completeness limit all the way to the highest mass bin.
At the high-mass end of the MS MF, saturation losses thin out the CMD.
As a consequence of the combination of different exposures taken under varying 
conditions, the saturation
limit is not a sharp point, but extends over a range of $\sim 1$ mag in $J_s$. 
The approximate fraction of stars lost due to
saturation in the upper mass bin, $M < 20\,M_\odot$, is $\sim 50\%$. 
The high-mass end of the MF has been
corrected accordingly, and is included in the MF fit.
The MF in NGC\,3603\,YC is consistent 
with a single-slope power law over the entire mass range probed by 
our data, $0.4 < M < 20\,M_\odot$. Note that the apparent turn-over
below $0.4\,M_\odot$ is not included in the fit, and we are not 
able to draw conclusions on the possible turn-over mass in NGC\,3603\,YC,
because the MF is heavily depleted by incompleteness, and the correction
increasingly uncertain.

All resultant MF slopes are summarised in Tab.~\ref{n3603mftab} and sample MFs are shown
in Fig.~\ref{mf_all}. 
The derived individual slopes of the MS alone (see Fig.~\ref{mf_all}
and Tab.~\ref{n3603mftab}) are meant 
to illustrate effects of the different corrections introduced, and should not
be taken at face value due to the low number of bins included in these fits. 
The PMS isochrone covers a large range in color over a narrow
mass range in the transition region where PMS stars equilibrate to main sequence hydrogen burning.
The colors in the transition region range from $J-K=0.7$ mag to $J-K=1.3$ mag
with masses between 2.5 and $4\,M_\odot$. Only four model points are available 
for the transition region, such that the mass transformation had to rely on 
interpolation between theoretically predicted color, magnitude and thus mass
values. This is likely the cause for the enhancement of stars seen in each MF
in the mass bin centered on $2\,M_\odot$, and the subsequent depletion in the number of
stars in the following mass bin centered on $3.2\,M_\odot$.
The combined MS/PMS mass function is less affected by the two subsequent
bins at PMS/MS transition masses, such that the continuous mass function yields the
most reliable fit. 

After the effort undertaken to correct for the effects discussed above, 
the derived mass function slopes show remarkably small scatter. 
A few general tendencies are observed. First of all, the slope of the 
PMS MF is very stable with values ranging from $\Gamma=-0.61$ to $-0.85$, with a scatter 
smaller than the formal least-squares fitting error of $\sim 0.25$. 
The MS MF is more severely affected by binary correction than the PMS MF. 
This is expected as the upper mass bins are emptied without substitution
from higher mass bins due to the saturation truncation, while binary intermediate- and 
low-mass stars are replenished from brighter bins. As the upper end of the MF
is heavily affected by saturation, the loss in stars and resultant steepening of the MF
is a consequence of saturating the {\sl binary} sequence. 

The slope of the combined mass function is not considerably affected by the MS variations,
but shows remarkable consistency with values ranging from -0.84 to -0.99 with a formal fitting 
error of $\pm 0.1$. As expected for the core region with $R < 1$ pc, field contamination 
has only a minor effect on the MF slope (Tab.~\ref{n3603mftab}, cf.~Fig.~\ref{density}).
The use of an older main sequence of 2 or 3 Myr does not change the combined
MF fit significantly (see Tab.~\ref{n3603mftab}), as expected\footnote{Note that for the PMS
only the best-fitting model was used, while other models and ages unable to reproduce
the stellar population in the CMD are not considered (see Andersen 2004 for a discussion 
on MF slope variations as a function of PMS models).}.
The most reliable derivation of the mass function in the central cluster area of NGC\,3603\,YC
including all corrections discussed in the previous sections is displayed in the bottom panel
of Fig.~\ref{mf_all}.
The slope of the  mass function including field subtraction,
individual dereddening and binary correction is $\Gamma=-0.91 \pm 0.08$
over the mass range $0.4 < M < 20\,M_\odot$.

A slope of $\Gamma = -0.77$ down
to $M > 1\,M_\odot$ was derived by Eisenhauer et al.~(1998) from high-resolution ADONIS 
data in the innermost 13$^{''}$ of the cluster center, in reasonable agreement with
our estimate. These authors did not account for binaries, and interpreted the 
observed scatter in the PMS/MS transition as an age spread in the range 0.3 to 1 Myr. 
Accordingly, a mean age of 0.5 Myr was used to derive masses of PMS stars. 
As shown in Paper I, Fig.~4, this interpretation is consistent with the data
as long as the photometric accuracy does not allow one to resolve the equal-mass
binary offset. A younger age and thus evolutionary state results in assigning
lower masses to the stars. Comparably, binarity correction results in assigning stars 
with a given brightness a (well-defined) lower mass. As the 0.3 Myr isochrone coincides
well with the binary sequence, the masses on the PMS derived for the transition 
stars are comparable. On the other hand, the effect of possible companions
on the MF is not taken into account in the case of an age spread. Statistically
adding companions at a constant fraction in all magnitude bins should, however, 
not alter the MF slope. 
Thus a comparable slope in the MF using a younger isochrone for the transformation of
magnitudes into masses as opposed to binary correction can be understood from 
the similarity of the 0.3 Myr and the binary isochrone. 

A severe difference between both data sets remains, limiting the above comparison. 
Mass segregation in the cluster, in the sense that massive stars may be found 
predominantly in the cluster center as observed in the case of the Arches cluster \citep{Stolte2003},
would cause the MF slope to steepen with increasing radius. As the ISAAC data 
are limited to $R > 7^{''}$ due to saturation, while the ADONIS data 
only extend out to $R < 13^{''}$, a steeper MF would be expected from ISAAC
than from the central ADONIS data in the presence of mass segregation. 
The comparison is complicated by the different procedures used. When the MF 
is calculated with mere field subtraction, without dereddening and binary correction
for a 1 Myr isochrone, the slope becomes $\Gamma=-0.86$, slightly, but not 
significantly steeper than the ADONIS slope. No significant change can be found
between the ADONIS core MF and the ISAAC MF outside the core. This agreement is 
even more surprising as the core is known to be mass segregated within $R < 6^{''}$
from HST/WFPC2 data \citep{Sung2004}. Sung \& Bessel derive a MF slope on the HST/PC
area of $40^{''}\times 40^{''}$ ($R \lesssim 20^{''}$) of $\Gamma = -0.9 \pm 0.1$
in the {\sl initial} mass range $1 < M < 100\,M_\odot$.
This slope agrees surprisingly well with our slope using present-day masses.
Stellar evolution at these young ages only becomes significant for stars
in excess of $\sim 50\,M_\odot$, such that the mass range of $0.1 < M < 20  \,M_\odot$
covered by our ISAAC data is not influenced by post-main sequence evolution.
The correspondence in both MFs might be a cause of the consistent age 
estimate of a 1 Myr single-age starburst in both derivations, but the consistency 
is still surprising in view of the different procedures and evolutionary models 
applied. In summary, the integrated MF slope in NGC\,3603\,YC is with $-0.9$ slightly
flattened with respect to a Salpeter slope, $\Gamma=-1.35$, suggesting the efficient 
formation of 
high-mass stars in the cluster center, but not a strong deviation from a normal IMF.
%
%
%

\section{Radial variation in the mass function of NGC\,3603\,YC}
\label{mfradsec}

Radial mass functions were derived including field subtraction,
individual dereddening and correction for binaries on the MS as well as 
on the PMS as described above. Three annuli covering equal areas were 
selected, each covering 1/3 of the area enclosed within $7^{''} < R < 33^{''}$.
The smallest accessible radius of $7^{''}$ is determined by saturation 
in the cluster core.
The radial selections correspond to $7^{''} < R < 20^{''}$, 
$20^{''} < R < 27^{''}$, and $27^{''} < R < 33^{''}$ (cf. Fig.~\ref{clustersel}).
The mass functions for these annuli are shown in Fig.~\ref{n3603mfrad}.

The slopes of the MFs displayed in Fig.~\ref{n3603mfrad} in the innermost two annuli
agree roughly within the uncertainties, and only a marginal increase from $-0.9 \pm 0.15$ 
($7 < R < 20^{''}$) to $-1.1\pm 0.15$ ($20 < R < 27^{''}$) is observed. 
Sung \& Bessel (2004) observe an increase from $\Gamma=-0.5 \pm 0.1$ 
in the cluster core ($R < 6^{''}$) to $-0.8 \pm 0.2$ for $6 < R < 12^{''}$ 
and $-1.2 \pm 0.2$ for $12 < R < 20^{''}$, suggesting that mass segregation 
is most pronounced in the core and the MF steepens to normal values rapidly
beyond the core radius, in good agreement with our values for $\Gamma$ taking 
into account the different radial selections applied.
In the outermost annulus, $27 < R < 33^{''}$, the MS is almost entirely depleted,
and a reasonable power law fit can only be applied to the PMS.
The PMS fit yields $\Gamma=-0.8 \pm 0.2$, such that a further systematic 
increase in the slope at larger radii is not supported by our data. 
 
Despite the comparable slopes in all annuli, the truncation
and depletion of high-mass stars yields strong evidence for mass segregation
in NGC\,3603\,YC. This indicates that the slope alone is not adequate 
to describe mass segregation processes.
While the low-mass (PMS) distribution decreases in number as the 
density decreases with larger radial distance, the shape in each 
annulus shows a close resemblance. The high-mass (main sequence) end 
on the other hand becomes more and more depleted with increasing radius. 
From the incompleteness corrected number counts, we can estimate 
whether the MS depletion is consistent with the expected decrease 
in stellar density as observed for the PMS.
While the cluster center ($7^{''} < R < 20^{''}$) harbors $\sim 250$ PMS
stars, the PMS population in the second annulus decreases to $\sim 87$
or 35\% of the central population. In the third annulus, 72 PMS stars remain, 
indicating a slow decrease in the density of 
low-/intermediate-mass stars. The MS population harbors 78 high-mass stars
in the innermost annulus. With the same decrease in stellar density, 
27 MS stars are expected in the second annulus, while only 17 are observed,
and 23 should still be present in the third annulus, where only 4 (!) MS stars
are found. We can conclude that the depletion on the MS is much stronger 
than expected from the decrease in stellar density with increasing distance 
from the cluster center, providing strong evidence for segregation of 
massive stars inside the cluster core.

In stark contrast to the strong outwards variation in the MS
population, the shape of the PMS distribution of stars remains similar
throughout all annuli. The truncation at very low masses in the 
first and second annulus is due to crowding losses of
faint stars, as indicated by the incompleteness correction.
As the lowest mass bins are incomplete by more than 90\%, 
a final conclusion on a possible change in the low-mass population 
can only be drawn for stars more massive than $1\,M_\odot$, where the completeness
fraction is more than 50\% in all annuli. No significant change is observed
in the 3 mass bins above this threshold. The 50\% limit progresses down
to $0.4\,M_\odot$ in the two outer annuli, where no strong variation
in the MF is observed between both these annuli. Even when the analysis is 
extended to radii of $65^{''}$, well outside the distinct cluster center,
the field subtracted MF displays the same slope in the PMS regime (see Tab.
\ref{n3603mfradtab}). The relative fraction of stars
detected down to the lowest mass included in the Palla \& Stahler models,
$0.1\,M_\odot$, increases noticeably in the outermost annulus. 
Taking into account the incompleteness correction in all annuli,
a truncation in the mass function of NGC\,3603\,YC at the low-mass end
can safely be ruled out down to $0.4\,M_\odot$.

\section{Mass segregation in NGC\,3603\,YC}
\label{massseg}

As the slope of the combined MS+PMS mass function does not vary 
significantly with radial distance from the cluster center, this measure 
cannot be used to draw conclusions on mass segregation in NGC\,3603\,YC.
From the mass functions, there is strong evidence 
that the PMS population alone, covering a mass range of $0.4 < M < 4\,M_\odot$
from the 50\% completeness limit to the PMS/MS transition, is 
not segregated. There is no indication of a flattening or depletion
of the MF at the low-mass end towards the cluster center.
Nevertheless, the truncation of high-mass MS stars with $M > 4\,M_\odot$
beyond a radius of $27^{''}$ is a clear indication of mass segregation
in the {\sl massive} population. Such a behaviour can be expected
from primordial mass segregation assuming that high-mass stars form
predominantly in the densest cluster region, and may thus have 
resided close to the cluster center throughout their lifetime, while 
low-mass stars can form anywhere inside the parental cloud. However,
dynamical segregation will also transport high-mass stars inwards
and low-mass stars outwards during interactions.

Although in a cluster as young as $\sim 1$ Myr a location of 
massive stars in the cluster center is very suggestive of primordial
segregation, Bonnell \& Davies (1998) derive a similar stellar 
distribution from N-body simulations of dynamical cluster evolution. 
For their richest cluster with 
1500 stars statistically drawn from a combined Salpeter (1955) and 
Kroupa et al.~(1990) MF truncated at $m_{low}=0.1\,M_\odot$, 
and a standard number density profile, $n \propto r^{-2}$,
90\% of the massive stars migrate into the innermost 0.5 half-mass
radii within 10 crossing times, no matter whether the massive
stars were originally randomly distributed in the cluster potential
or concentrated near the half-mass radius. Independently of the
exact segregation timescale, these models thus also yield strong 
evidence that initial conditions are lost rapidly.
Although we cannot derive the core radius and in particular the
crucial contribution of the massive component in the core to the cluster
potential from the ISAAC data, a core radius of $\sim 0.2$ pc is estimated
from HST/WFPC2 observations (Grebel et al., private communication). 
Furthermore, a velocity dispersion 
of a few km/s is typically observed in dense clusters. This core radius
leads to a crossing time of only $10^5$ yr in the case of a velocity dispersion of 
2 km/s. According to Bonnell \& Davies, heavy mass segregation should
have created a high-mass dominated cluster core after $10 t_{cross}$ or in our 
example only 1 Myr. This order of magnitude estimate suggests that the segregation 
timescale expected from cluster evolution models is comparable to the 
cluster age, and thus that the cluster was already shaped by heavy dynamical segregation.
This early loss of initial conditions prohibits conclusions on the formation locus
of high-mass stars. We can therefore not distinguish primordial from dynamical 
segregation in NGC\,3603\,YC from the present-day distribution of high-mass stars. 

To derive more quantitative conclusions on the mass segregation in NGC\,3603\,YC,
two other means are investigated in the following sections.
First, cumulative functions were created from the mass distribution
to avoid the binning dependence inherent to the standard MF derivation.
Secondly, the number ratio of high-mass to low-mass stars is analyzed
as a function of the radius.

\subsection{Cumulative Functions}

Cumulative functions (CFs) are created by consecutively adding
stars with decreasing mass, starting with one star at the highest
mass observed. 
The field-subtracted, color-selected distribution of stars used 
to derive the mass functions enters the 
cumulative functions. Masses are assigned to individual stars
during individual dereddening in the same way as for the MF derivation.
The only difference to the mass distributions in the MFs
is that the artificially added companion stars for binary 
correction are not counted as stars in the CF. These ``stars''
have not been observed as individual sources, and the 
correction is - in particular on the PMS - purely statistical.
As the only indication for binaries is given by the binary 
candidate sequence observed in the higher mass population, 
only equal-mass binaries can be taken into account. 
Systems with smaller mass ratios cannot be distinguished 
from single stars. If the binary stars are not biased
to a certain mass range, the mass distribution in 
binary stars is the same as in single objects. Adding 
companion stars thus adds to the total mass, but not 
to the statistical properties of the distribution.
The brightness correction of $0.75$ mag for stars on the secondary sequence, 
however, is crucial, as $\sim 30\%$ of the individual masses are overestimated 
if the brightness is indeed enhanced in these stars (by binarity or other physical 
mechanims). Thus, the CFs are created
from all stars selected as cluster members in the CMD after 
field-subtraction, with masses corrected for the derived
binary fraction of $30\%$. 

%
The resultant cumulative functions of the central cluster
population as well as the radial variations in the CFs are 
shown in Fig.~\ref{mcum3603}. Theoretical CFs corresponding to a 
single exponent power-law MF are underlaid for exponents of
$-0.3 > \Gamma > -1.3$. For the main cluster area, $7^{''} < R < 33^{''}$,
the main sequence stars in the mass range $4 < M < 12\,M_\odot$
follow closely the $\Gamma=-0.7$ line. The irregular shape at 
higher masses is partially due to random sampling of the upper MF
as well as saturation losses for $M > 15\,M_\odot$. Between 3 and $4\,M_\odot$,
the irregularity arises in the ambiguity of stellar locations in the 
CMD and thus individual masses in the PMS/MS transition region.
The pre-main sequence distribution favours 
a flatter value of $\Gamma \sim -0.5$. 

In the cluster center ($R < 20^{''}$), 
the CF tangents the $\Gamma=-0.3$ curve for $M < 2\,M_\odot$, 
and $\Gamma$ tends to -0.5 at higher masses. At larger radii,
the CF falls towards steeper values of $\Gamma$ as expected 
in the case of mass segregation. On the PMS, however, this effect is weak, and the
general shape of the CF remains remarkably constant out to radii where almost
no main sequence stars remain. 
This is particularly pronounced in Fig.~\ref{mcumall3603},
where the PMS region below $3\,M_\odot$ can hardly be distinguished
in all annuli except for the slightly flatter CF in the cluster center.
This supports the finding of N\"urnberger \& Petr-Gotzens 
(2002) of a cluster extent of $R\sim 150^{''}$ beyond the edges
of our field of view. This pre-main sequence population can 
be interpreted as an extended halo of low-mass stars comparable
to the halo observed around the Trapezium in the Orion Nebula Cluster 
\citep{Hille1997}.

The situation at the high-mass end is very different. For stars with 
masses $M > 2\,M_\odot$, the CF in the outer annuli indicates 
increasingly steeper values of $\Gamma\sim -0.7$ to $-0.9$. 
At larger radii, the high-mass end becomes increasingly depleted, 
evidenced in a steep decline of the CF toward higher masses. 
The most striking feature in the main sequence population is the sharp
truncation of the CF at $\sim 2.5\,M_\odot$ for radii $R > 27^{''}$.
This cut-off corresponds to the absence of main sequence stars in 
the mass functions (cf. Fig.~\ref{n3603mfrad}), and indicates that 
massive stars are highly segregated toward the cluster center. 
Beyond $33^{''}$,
the irregular CF shape and increase in the scaled frequency of 
high-mass stars, obvious in Fig.~\ref{mcumall3603}, marks 
the onset of field contamination, as at large radii statistics 
of cluster stars become poorer and realistic field subtraction more 
difficult than in the dense cluster center where cluster members dominate.

From these cumulative distributions, we see that
binning effects influence the slope of the mass functions 
significantly, and the assumption of a single exponent power-law 
distribution is over-simplified. While the MF concept is 
useful as a tool to compare results from different data sets
and regions, yielding a single characteristic value, the
assumptions entering such analyses have to be carefully
reviewed before conclusions on ``universality'' or ``deviation''
from a certain mass distribution are drawn.

\subsection{Fraction of high- to low-mass stars}
\label{fracsec}

The fraction of high-mass main sequence stars to low-mass pre-main 
sequence stars, $f (high/low) = n(> 4\,M_\odot)/n(< 4\,M_\odot$),
is derived as a function of distance from the cluster center. 
Individual mass derivations for stars in each separate annulus 
are used including 
field-star and incompleteness corrections for each radial
selection. In addition, the cluster population within
$7^{''} < R < 33^{''}$, yielding better statistics than the outer annuli,
is split into $5^{''}$ bins in steps of $2.\!^{''}5$ to 
allow a more complete radial coverage. The MS/PMS transition
mass, $M = 4\,M_\odot$, as the natural mass separation in the
stellar population of NGC\,3603\,YC is chosen to distinguish high- and 
low mass stars. Consequently, the fraction
of high- to low-mass stars also reflects the fraction of MS
to PMS stars.

%
A strong decrease in the fraction of massive vs. low-mass
stars is seen in Fig.~\ref{frac}. While 44\,\% of 
all stars with $M > 1\,M_\odot$ in the cluster core, 
$7^{''} < R < 10^{''}$,
have masses above $4\,M_\odot$ (and $M < 20\,M_\odot$), 
this fraction drops to 
20\,\% already at a radius of $20^{''}$. At a radius of $33^{''}$,
the contribution of massive stars is below 5\,\%. Beyond this radius,
field contamination at the bright end skews the ratios
again towards higher values. The steep decrease
reflects the high-density concentration of massive stars in the
cluster center, while at larger radii, $R > 30^{''}$, the cluster
is dominated by the extended low-mass population. 

An even steeper drop is observed in the {\sl mass fraction}
of high- vs. low-mass stars (right panel in Fig.~\ref{frac}).
While the total mass in high-mass stars exceeds the low-mass
contribution by a factor of 1.8 close to the cluster center,
meaning that high-mass stars contribute 64\,\% to the total mass,
the ratio of the mass in high- vs. low-mass stars drops to 
0.2 at $R=33^{''}$, such that high-mass stars contribute
only $\sim 17\,\%$ of the total mass at large radii while 
dominating the mass in the cluster core. Note that in this 
estimate, the compact, massive core of the starburst is excluded
due to saturation. The flat core MF ($\Gamma=-0.5$) derived by 
Sung \& Bessel (2004) for radii $R < 6^{''}$ indicates a strong
bias to high-mass stars in the core, such that including the 
massive core population would enhance the percentage of mass 
contributed by high-mass stars in the center of NGC\,3603\,YC.

As can be seen in Fig.~\ref{frac}, these variations in the 
number and mass fractions with radius are a very efficient 
tool to quantify mass segregation effects.

\section{From Orion to R\,136 - a structural comparison}
\label{comparesec}

Starburst clusters as massive and compact as NGC\,3603\,YC are
rare in our close neighborhood. In the Milky Way, 
only the Arches cluster close to the Galactic Center is known 
to have comparable mass and density \citep{Stolte2003}. 
Clark et al.~2005 argue that the recently unveiled Westerlund 1 
cluster is comparable to Arches in core density and might have
a total mass of $10^5\,M_\odot$ under the assumption of a 
normal IMF, but only the most massive stars down to type B0 
are analysed so far, while the low-mass population is 
as yet unknown.
In the literature, NGC\,3603 is extensively compared to the massive 30 Doradus 
star-forming complex in the Large Magellanic Cloud (LMC), hosting the 
dense cluster surrounding the compact, star-like source R\,136,
which features a core density derived from O and WR stars 
comparable to the central NGC\,3603 starburst cluster (e.g., Moffat 1994).
With a total mass of $2\cdot 10^4\,M_\odot$, the central
cluster around R\,136 in 30 Dor is at the upper mass end of local
young star clusters, and at the same time at the low-mass
end of Milky Way globular clusters. The large distance of 
6 and 8 kpc to NGC\,3603\,YC and Arches, respectively, 
allow only limited study of their low-mass population. 
This limitation applies to R\,136 and its surrounding cluster
at 50 kpc even more severely.
Because of its proximity to the Sun, the Orion molecular
cloud, on the other hand, is one of the best studied star-forming 
regions in the Milky Way. At a distance of only 450 pc,
the Orion nebula is sufficiently close to resolve the embedded 
population down to the substellar regime \citep{Muench2002}.
The extended molecular cloud hosts the young Orion Nebula Cluster
(ONC) containing stars with ages between 0.3 and 1 Myr in its center.
The Trapezium system, a dense cluster of massive stars, forms the core 
of the ONC. It is particularly interesting for the comparison with NGC\,3603 
and Arches that the ONC provided a spatially well-resolved testcase where mass 
segregation could be analysed quantitatively (Hillenbrand 1997). 
The Trapezium cluster is with a core density 
and mass of about one order of magnitude lower than NGC\,3603\,YC 
at the low-mass end of {\sl massive} star-forming regions in the Milky Way
(see Tab.~\ref{comparetab}),
albeit an example of the mode of star formation we understand as
normal in our local environment. As these star-forming regions play 
a major r\^ole in our understanding of the stellar IMF, we wish to 
compare their characteristics in the context of the influence of 
the star-forming environment on cluster formation.
Tab.~\ref{comparetab} summarizes the known properties of these
massive star-forming regions and their central clusters.
Properties of the Antennae starburst clusters and Milky Way globular 
clusters are included for reference. 
%

\subsection{Comparison of cluster characteristics}

\subsubsection{NGC\,3603}

The extent of NGC\,3603\,YC is estimated to a radius of 4.4 pc 
by N\"urnberger \& Petr-Gotzens (2002), beyond which it becomes
indistinguishable from the stellar field population. 
The core radius is estimated to 0.2 pc from a King-profile fit to 
the density profile (Grebel et al., private communication).
Within a radius of $\sim 1$ pc, a total mass of $7000\,M_\odot$ is  
derived via extrapolation of the observed MF down to $0.1\,M_\odot$ from 
a combination of HST/WFPC2 core data and the ISAAC observations presented above 
\citep{Stolte2003}. This estimate of the total cluster mass is a lower limit
as it does not correct for the low-mass halo, which is difficult to quantify
without independent field data. In the inner 0.12 pc ($d=6$kpc) of the dense 
cluster core, 
six O3 stars and three WN6 components are identified by Drissen et al.~1995. 

\subsubsection{R\,136 and its surrounding cluster}

As a consequence of the similarly compact central cluster and stellar mass 
content in the inner 1 pc, NGC\,3603 was suggested to be 
``a clone of R\,136'' by Moffat et al.~1994. 
Both clusters are with an age of 1-2 Myr equally young,
and contain a massive O-star population in their centers.
With $2\times 10^4\,M_\odot$ \citep{Hunter1995},
the total mass in stars in the cluster around R\,136 
is a factor of 2-3 larger than in NGC\,3603\,YC. 
With the given stellar mass content, the central 30 Dor cluster 
was very productive in forming high-mass stars.
Within the inner 0.2 pc, six O3 stars are classified by 
Massey \& Hunter (1998), comparable to the six O3 stars confined 
to the core of NGC\,3603\,YC ($r < 0.12$ pc). 
Despite this similarity in the immediate core, 
Massey \& Hunter (1998) classify
65 stars spectroscopically with types earlier than B0 out to 
radii of 5.7 pc,
of which 40 stars are of type O3, 4 are WN6 stars,
6 are in the O3/WN6 transition phase to Wolf-Rayet stars, 
and one O3-O3 binary is found, adding up to a total of 52 
early O stars. In the 65 brightest stars with types B0 and earlier
spectroscopically analyzed by Massey \& Hunter,
a strong bias to O3 stars is observed. A comparably
large number of O3 stars is not known in any other local
star-forming environment. Out of these, 21 O3 objects
reside within 1 pc from R\,136, while
no more O3/WN6 sources except for the nine very early 
type stars in the innermost core of NGC\,3603\,YC are detected 
in the central 1 pc covered by the ISAAC data and further out
to 4.2 pc (Moffat et al.~1994), the suggested cluster extent \citep{Nuern2002a}.
In NGC\,3603\,YC, 84\% (31) of the early O-star population 
($M_V > -5.0$, $M \sim 25\,M_\odot$) is confined to the innermost 
0.5 pc, and all O-stars are found within 1.15 pc from 
the cluster center. A comparable number of 33 O and WN stars
is classified by Massey \& Hunter (1998) in the inner 1.1 pc of 
the cluster surrounding R\,136.
However, this comprises only 63\% of the 52 most luminous stars
in the inner 30 Dor cluster down to the same $M_V$ limit, 
while the remainder is spread out to radii as large as 4.2 pc 
(see Moffat et al.~1994, Fig.~9).

Comparable to the morphology observed in NGC\,3603\,YC, the cluster 
around R\,136 displays a distinct dense cluster core surrounded 
by a stellar halo. The main difference between NGC\,3603\,YC 
and the core of 30 Dor is given by
the spatial extent of the halo and its content of massive stars.
While the dense cluster center around R\,136 is with a radius of 4.7 pc 
\citep{Hunter1995} comparable to the {\sl total} cluster radius of 4.4 pc 
derived for NGC\,3603\,YC by N\"urnberger \& Petr-Gotzens (2002), 
a halo of young, massive stars with Wolf-Rayet characteristics
is detectable out to a distance as far as 130 pc from R\,136, 
beyond which it merges smoothly into the field 
population \citep{Moffat1987}. From the fact that only two more 
WNL stars are found in the rest of the LMC, Moffat et al.~1987 argue 
that the observed young population of WN6/7 stars are members
of 30 Dor. If R\,136 and 30 Dor were formed in a single starburst, 
this renders the halo around R\,136 a factor of 30 larger 
than the radial extent of NGC\,3603\,YC. 

\subsubsection{Trapezium}

The ONC contains $\sim 2000\,M_\odot$ within a radius of 2.5 pc
\citep{Hille1997}.
The ONC appears as an extended halo of low-mass stars around the 
dense central Trapezium cluster. The core radius in the cluster 
derived from King model fitting is 0.2 pc (Hillenbrand \&
Hartmann 1998), identical to NGC\,3603\,YC. The central 
{\sl stellar number} density inside the core is also very comparable,
with $\rho_c=1.7\cdot 10^4\, {\rm stars/pc^3}$ in the Trapezium
core and $\rho_c=1.4\cdot 10^4\, {\rm stars/pc^3}$ in NGC\,3603\,YC.
There is, however, one significant difference between both 
clusters - the central {\sl mass} density is with 
$\sim 10^5\,M_\odot/{\rm pc^3}$ (Moffat et al.~1994) in the starburst about one 
order of magnitude higher than the $2\cdot 10^4\,M_\odot/{\rm pc^3}$ in
the Trapezium core. Given the large number of at least 40 O-stars 
in NGC\,3603\,YC (Drissen et al. 1995), versus a mere
$\sim 10$ OB stars in the ONC, the difference in mass density is not 
surprising. NGC\,3603\,YC contains the entire mass found in the ONC
{\sl only} within its core. 

\subsubsection{Arches}

The Galactic Center Arches cluster falls with a total mass of $\sim 10^4\,M_\odot$
and central density of $3\cdot 10^5\,M_\odot {\rm pc^{-3}}$ inbetween NGC\,3603\,YC
and the cluster surrounding R\,136. 
As such, it is the most massive, young compact cluster in the Milky Way
for which a detailed MF analysis is available. 
The core radius is estimated to be $\lesssim 0.24$ pc from the half-mass 
radius \citep{Stolte2003} comparable to both NGC\,3603\,YC and Trapezium, although 
the density profile cannot be used to derive a fitted core radius 
because it is heavily distorted by tidal disruption in the Galactic 
Center gravitational field. Due to the high stellar field density in the GC environment, 
the true extent and the population surrounding the cluster beyond a radius of 1 pc is
currently not known. A part of the cluster population has likely formed an extended
tail or halo, as low- and intermediate-mass members are stripped off the cluster rapidly 
in the GC tidal field. Nevertheless, a compact, highly mass segregated high-mass core 
is observed in the center of a more extended intermediate-mass population, in structure
comparable to NGC\,3603\,YC, but with a larger characteristic mass scale 
\citep{Stolte2005, Stolte2003}.

\subsection{MF slopes in four clusters}

The central 4.7 pc of R\,136/30 Dor exhibit a normal mass function
with a slope of $\Gamma=-1.3 \pm 0.1$ in the mass range
$15 < M < 120\,M_\odot$, identical to the slope of a Salpeter IMF.
The massive end, $30 < M < 120\,M_\odot$, 
of this MF is obtained from spectroscopy of the high-mass 
population (Massey \& Hunter 1998). In the MF derivation, 
the $120\,M_\odot$ upper mass limit is imposed from the truncation
of stellar evolution models. From luminosity and - where available -
spectroscopy of the brightest, highest-mass stars, Massey \& Hunter
suggest initial masses of up to $150\,M_\odot$. Given the total mass
of the cluster, these authors argue that this is in agreement with an 
untruncated  normal mass function up to the highest masses.

Hunter et al.~(1996) study the radial variation in the MF slope 
in the mass range $2.8 < M < 15\,M_\odot$ in four annuli with
$0.11 < R < 1.1$ pc using HST/PC data. From inner to outer annuli, values
of $-1.0 \pm 0.4$, $-0.7 \pm 0.4$, $-1.1 \pm 0.3$ and 
$-1.05 \pm 0.2$\footnote{cf. Massey \& Hunter (1998) for corrections 
of the values given in Hunter et al.~(1996)}, respectively,
are derived, showing no variation with radius within the uncertainties.
The same behaviour was derived in the case of NGC\,3603\,YC in 
Sec.~\ref{mfradsec}. The HST/PC data analysed by Hunter et al.~1995 
only resolve the cluster beyond 0.1 pc. Mass segregation in the form of a 
flattened MF slope is observed in NGC\,3603\,YC only inside the core radius 
of 0.2 pc \citep{Sung2004}, while segregation at larger radii is evidenced 
in the decrease in the mass fraction in high-mass stars (Sec.~\ref{fracsec}) 
and the truncation of the MF at the high-mass end
(Sec.~\ref{mfradsec}), but not in the slope of the MF. 
A mass segregated core biased to high-mass stars can thus not be entirely 
ruled out for R\,136, although the HST/PC image presented in Massey \& Hunter
1998 (see their Fig.~2) suggestes that the center of R\,136 is well 
resolved with HST, and the MFs in Hunter et al.~1996 do not show
a depletion at the high-mass end as seen in NGC\,3603\,YC. 
The mass function slope of $\Gamma=-1.0$ to $-1.1$ ($2.8 < M < 15\,M_\odot$) 
observed in approximately the same radial range, $0.1 < R < 1.1$ pc, agrees 
very well with the slope of $\Gamma=-0.9 \pm 0.1$ ($0.4 < M < 20\,M_\odot$ in 
$0.2 < R < 1.0$ pc) in NGC\,3603\,YC, both subject to a similar saturation 
limitation at the massive end. 

The mass function in the Trapezium system was investigated most recently
by Muench et al.~(2002). A slope of $\Gamma = -1.2 \pm 0.2$ is 
found for $0.6 < M < 5\,M_\odot$, consistent with $\Gamma \sim -1.3$
found by Hillenbrand (1997) over a mass range $0.25 < M < 12\,M_\odot$,
the upper limit of the ONC IMF power law tail.
These slopes are comparable to the overall slope in R\,136 and its 
surrounding cluster ($2.8 < M < 120\,M_\odot$, Massey \& Hunter 1998) 
despite the different mass ranges covered,
supporting the idea of a universal power-law IMF above the turn-over mass.
Again, no indication of a radially increasing slope is found outside the 
cluster core \citep{Muench2002}. 

In the central parsec of the Arches cluster, the same slope of $\Gamma = -0.9 \pm 0.15$ 
as observed in NGC\,3603\,YC is obtained for the intermediate-
to high-mass regime, $4 < M < 65\,M_\odot$. 
The low-mass regime of the cluster is not yet 
resolved. Although all of the above slopes barely 
agree within the errors, the low fitting uncertainties probably
underestimate the true slope uncertainty, as photometric
age or distance uncertainties are not taken into account.
Thus, there may be a weak indication of a flattened mass 
distribution in NGC\,3603\,YC and Arches, while the MFs in 
the clusters surrounding R\,136 and the Trapezium follow a 
standard Salpeter law. 

Although the spatial distribution of high-mass stars around R\,136 
differs drastically from the morphology observed in Trapezium, 
NGC\,3603\,YC and Arches, 
the slope of the mass function is almost identical.
As already discussed in context of the radial variation
and mass segregation in NGC\,3603\,YC, this supports the conclusion
that the slope of the MF alone is not sufficient to understand
mass segregation.

Given the diversity of the star-forming environments considered, 
the slopes of the present-day MFs are in remarkable agreement,
in accordance with a universal IMF slope. 

\subsection{Primordial vs. dynamical segregation?}

In all four clusters, a heavily mass segregated core is observed.
Mass segregation in young clusters
was shown for the ONC and Trapezium in Hillenbrand 1997 (see also
Hillenbrand \& Hartman 1998).
Hunter et al.~1995 conclude from dynamical considerations that the 
core inside R\,136/30 Dor is entirely mass segregated for radii $r < 0.4$ pc,
and their IMF suggests that at least 40\,\% of the total mass are confined
within this radius. We estimate a half-mass radius of 0.24 pc for Arches, thus
about 50\,\% of the mass are confined within the core, 
similarly indicated for NGC\,3603\,YC from the HST analysis \citep{Stolte2003,Sung2004}.
Do these heavily high-mass biased, mass segregated cores imply a preferential 
formation of high-mass stars in the cloud cores, and thus primordial segregation?

Although we are not in the position to give a final answer to this question, 
there are several indications we can derive from the comparison of the known 
clusters. In all four clusters, the present-day MF rapidly approaches a slope 
close to a standard IMF outside the core radius. In NGC\,3603\,YC we have seen 
that segregation mostly affects the high-mass stars, while the shape of the low-mass 
distribution appears constant with radius. A similar behaviour is observed in 
the form of a constant MF slope in the cluster around R\,136 and Trapezium. 
In the Arches cluster 
the case is not clear due to the severe crowding and field contamination at masses
below $\sim 4\,M_\odot$. Despite these similarities in the intermediate-mass MF, 
NGC\,3603\,YC and 30 Dor show significant structural differences in the spatial 
distribution of high-mass stars. While both NGC\,3603\,YC and Trapezium 
exhibit a halo of low-mass objects, the cluster surrounding R\,136 is the only 
local starburst cluster that exhibits an extended halo of {\sl high-mass} stars. 
Hunter et al.~(1995) argue that the dynamical relaxation process is completed
around R\,136 in the innermost 0.4 pc, with the relaxation timescale
for high-mass stars being much smaller than the cluster age within this radius.
This situation resembles the derivations for NGC\,3603\,YC \citep{Stolte2003} and 
Orion \citep{Hille1998} in the respect that a short dynamical
relaxation time prohibits one to distinguish primordial and dynamical
segregation in the cluster center. In the extended halo surrounding R\,136, however, 
the relaxation timescale is much larger due to the lower density
and larger distance from the cluster center, such that stars formed
in the outer regions of the cluster are not significantly segregated,
but should still be found close to their birth positions.
This implies that even the formation of the highest mass stars is
{\sl not} confined to the densest cluster region, as observed in the two
Milky Way starburst clusters. 
Nevertheless, even the 
Milky Way hosts OB associations with lower stellar density capable of forming
numerous O stars. The most prominent example discovered so far is the Cyg OB2
complex, which harbours 48 O stars with one star classified as early as O3
(Comer\'{o}n et al.~2002, Hanson 2003). The young stellar population is extended
over an area of $\sim 20$ pc $\times 20$ pc and lacks a compact core as opposed 
to the 1 pc into which the high-mass core of NGC\,3603\,YC is confined. 
Furthermore, populations hidden
by extinction such as Westerlund 1 with an estimated core density of 
$\sim 10^5\,M_\odot {\rm pc^{-3}}$ are just recently unveiled by deep NIR 
observations. Westerlund 1 hosts at least 53 early-type stars and a total mass 
of $10^5\,M_\odot$ was suggested for the cluster (Clark et al.~2005), about one 
order of magnitude higher than estimated in NGC\,3603\,YC. Despite the 
comparable core density, O-type stars are found out to radii of 3\,pc from 
the cluster center in Westerlund 1, and the spatial distribution of massive
stars resembles the environment around R\,136 more closely than NGC\,3603\,YC. 
In which respects is 
the star-forming environment in the LMC 30 Dor complex different from 
the NGC\,3603 and Arches starburst environments? 

Several differences are observed between 30 Doradus and the NGC\,3603 molecular cloud.
First of all, the total gas mass in the 30 Dor complex is with $(1..6)\cdot 10^7\,M_\odot$ 
\citep[depending on the CO/H$_{\rm 2}$ conversion factor]{Cohen1988} 
at least one to two orders of magnitude larger than the $4\cdot 10^5\,M_\odot$ 
determined with the same method in NGC\,3603 \citep{Grabelsky1988}. 
The lower metallicity in the LMC is frequently employed as the origin for differences
in the young stellar population. However, the metallicity in 30 Dor has recently 
been shown to be 2/3 of the solar value \citep{Peimbert2003}, such that significant 
changes in the outcome of the star-forming process are unlikely due to the marginally
lower metallicity. Another significant difference was, however, suggested early:
Werner et al.~1978 estimated an optical path length of 30 pc in 30 Dor vs. 0.1 pc 
in Orion from the much lower far-infrared optical depth, indicating a low density of dust
and molecular material in the giant cloud. If photons travel longer distances, 
the heating of molecular cores may be much more efficient out to larger distances
from the dense cloud center, where the first high-mass stars are likely to have formed.
The suggested optical path length is on the order of 100 times larger than in 
the Orion cloud. Taking into account the quadratically decreasing radiation field 
intensity with distance, one expects the spatial extent of clusters in Orion 
to be about one order of magnitude smaller than in 30 Dor, consistent with 
the 2.5 pc radius of the ONC. 
If the gas/dust distribution in NGC\,3603 is similar to Orion, 
this could explain the $\sim 30$ times more extended halo of 
130 pc around R\,136 vs. 4.4 pc in NGC\,3603. Observations of core temperatures
in the 30 Dor complex as well as a comparative study of optical path lengths in 
Milky Way GMCs would clearly be very valuable to probe this suggestion.

A higher average cloud temperature is actually observed in the Galactic Center 
environment. Typical values of 70 K with a range of 30-200 K are measured in the 
GC \citep{Morris1996} vs. $\sim 20$ K observed in the cores of moderate 
star-forming environments. The intense radiation field in the GC may have a similar
effect on core heating as the longer optical path length suggested for 30 Dor.
As the path length in the GC is likely very small due to high gas and dust
densities, the less efficient core cooling may locally cause the formation of
cores with higher mass rather than extended formation of high-mass cores as 
in 30 Dor. A low cooling efficiency and intensive external heating may result
in a higher core temperature when the opacity limit is reached and fragmentation
ceases, implying a lower core density and thus higher mass needed for gravitational collapse.
A lower critical density at which fragmentation stops results in less 
fragmentation and a higher average core and final stellar mass according to 
recent numerical simulations \citep{Jappsen2005}. The present-day MF
in the Arches cluster displays indications for a high-mass turn-over around 
$6\,M_\odot$ followed by a possible depletion towards lower masses 
\citep{Stolte2005}. In contrast to 30 Dor, where high-mass stars form
at large distances from the cluster center, but the overall IMF appears normal,
an enhanced core temperature in the GC may have caused a locally top-heavy 
IMF in the Arches core. 

The Arches cluster is known to be tidally disrupted rapidly by the 
strong GC tidal field. Thus, for this cluster a skewed present-day MF may well 
have resulted from dynamical segregation alone. In 30 Dor the spatial distribution 
of massive stars is unlikely to be caused by dynamical evolution, as the dynamical 
timescales in the outskirts of the cluster (beyond $r=1$ pc) are larger than the cluster age
\citep{Hunter1995}. Although NGC\,3603 as a counterpart to the Arches cluster 
evolving in a moderate star-forming environment is estimated to be dynamically 
significantly mass segregated (see Sec.~\ref{massseg}), and will likely not survive on 
globular cluster timescales due to its compactness, the dynamical timescales in 
NGC\,3603\,YC are still at least one order of magnitude longer than in the GC environment
\citep{Stolte2003}. Thus, the differences between Arches and NGC\,3603\,YC
can be caused by dynamical evolution, while the structural discrepancies
between NGC\,3603 and the central 30 Dor cluster likely originate from 
intrinsic differences in the parental clouds.

Hillenbrand \& Hartmann (1998) performed a detailed dynamical
analysis of the ONC and its core, the Trapezium, and the mass
segregation studied by Hillenbrand (1997).  
The radial variation in the mass distribution
observed in NGC\,3603\,YC and the ONC is comparable despite
the fact that the starburst cluster hosts at least 40 O-stars
as opposed to only 3 O-stars in the Trapezium system.
Hillenbrand \& Hartmann derive cumulative mass distributions
in different annuli for several mass bins. These CFs show a different
shape for massive stars with $M > 5\,M_\odot$, while the three
mass bins between 0.3 and $5\,M_\odot$ all 
show a remarkably identical behaviour. In the cluster core
($R < 0.5$ pc), however, all four mass intervals follow 
each other closely, indicating strong segregation in the core,
producing a bias to massive stars with respect to the low-mass
population. The similarity over the wide range in masses suggests
a flat MF exclusively in the core of the ONC. At larger radii, 
mass segregation in the ONC is observed down to $1-2\,M_\odot$,
but not below these masses. This is surprisingly similar to NGC\,3603\,YC,
where the CFs of the PMS population up to masses of 
$\sim 3\,M_\odot$ are indistinguishable in different annuli,
while for larger masses strong changes are observed 
(see Fig.~\ref{mcumall3603}).

From the central stellar density, the number of stars in the core, 
the mean density and the measured velocity dispersion of 2.5 km/s,
Hillenbrand \& Hartmann derive a relaxation time of 6.5 Myr in the ONC.
This timescale is longer than the maximum age of $\sim 1$ Myr, suggesting 
that some amount of primordial segregation had to be present.
The same authors also note that if only the massive core with
$\sim 400\,M_\odot$ is taken into account, the higher mean mass
and stellar density causes a decrease to $t_{relax} \sim 0.6$ Myr,
comparable to the age of the young cluster population. 
This means that in the Trapezium, too, observational evidence
for primordial segregation is weak, as the massive stars may 
have migrated to the cluster center within the lifetime of the cluster.
Bonnell \& Davies (1998) perform detailed N-body simulations
of ONC-type clusters, and find that the mean mass in the core
as well as the ratio of high- to low-mass stars can only 
be recovered if the massive stars are initially placed 
close to the core, in agreement with primordial segregation.
Given the high stellar mass and density in the core of NGC\,3603\,YC,
some amount of primordial segregation has likely contributed
to shaping the cluster. 

A weak indication for primordial segregation is given by the low-mass MF. 
Both the MF as well as the CFs in NGC\,3603\,YC and Trapezium 
\citep{Hille1998}  
display increasing truncation at the high-mass end with increasing cluster center
distance, but the shape of the low-mass distribution remains remarkably constant.
From purely dynamical segregation a systematic increase in low-mass stars with 
increasing radius and a corresponding decrease in the apparent median 
(and characteristic) mass is predicted. 
While measuring the radially varying characteristic mass in the form of 
the turn-over in the MF in NGC\,3603\,YC has to await higher resolution 
observations, the mass fraction in low-mass stars increases with radius in 
NGC\,3603\,YC as shown in Fig.~\ref{frac} consistent with expectations 
from dynamical evolution models. 
A detailed comparison with numerical simulations is needed 
to determine whether both the shape of the low- and high-mass MF as well as the 
increasing fraction of low-mass stars can be explained by dynamical segregation 
alone, or whether and how much primordial segregation is needed to reproduce
the high-mass core along with the radially invariant properties observed 
in the low-mass halo.

In summary, although we cannot finally distinguish between primordial vs. dynamical 
segregation in the four clusters, we observed striking similarities between NGC\,3603\,YC
and the ONC/Trapezium system, while structural differences are very prominent 
between NGC\,3603 and the starburst surrounding R\,136 in 30 Dor as well as 
the Arches cluster in the GC environment. 
From these we conclude that the star-forming environment, despite
surprising similarities in the efficiency to produce the same relative fractions of
high- and low-mass stars and thus the same IMF slope, influences the structural properties
and thus the long-term survival of starburst clusters severely, and therefore also 
the potential emergence of globular clusters from starburst environments.

\section{Summary and Conclusions}
\label{sumsec}

From high-resolution near-infrared photometry, the present-day MF in 
the Milky Way starburst cluster NGC\,3603\,YC is derived. 
Different treatments of the field star contamination, individual extinction 
correction, and binary estimates are used to obtain the MF. We show that 
the slope of the MF is surprisingly robust against the derivation procedure.
A slope of $-0.91 \pm 0.15$ is found in the mass range $0.4 < M < 20\,M_\odot$ 
for the integrated MF, in excellent agreement
with the optically derived MF by Sung \& Bessell 2004, who obtain the same
slope in the mass range $1 < M < 100\,M_\odot$ despite entirely different 
procedures employed to calculate the MF. This slope is identical to the 
slope found in the GC Arches starburst cluster, and only slightly 
flattened with respect to a Salpeter MF ($\Gamma=-1.35$).

Radial variations are observed in the spatial distribution of stars, 
especially reproduced in the fraction of high- to low-mass stars and thus 
the characteristic mass at a given radius, and the depletion of massive
stars at the high-mass end of the MF. This radial variation is, however, 
not reproduced in the slope of the MF with values of -0.9, -1.1, -0.8, -0.9, -0.9
in five equal-area annuli from immediately outside the cluster core out to 
radii of 2 pc. Although the slope is frequently employed as the only 
parameter comparing the MFs in different star-forming environments, 
it may not be the most significant value to use. The fraction of low-mass
stars, the characteristic mass and the upper mass limit reflect mass
segregation more closely. 

A detailed comparison between NGC\,3603, 30 Dor/R\,136 in the LMC, Orion/Trapezium, 
and the Galactic Center Arches cluster suggests that a mass segregated core
with an extended stellar halo may be a common cluster structure in a diversity 
of environments. While the formation locus
of the highest-mass stars appears confined to the cores of these two 
Milky Way starburst clusters, with the surrounding halo being comprised mainly 
of low-mass stars, stars as early as spectral type O3 seem capable 
of forming at large radial distances from the cluster center around R\,136
in 30 Dor. Interestingly, the morphology in the Galactic star-forming 
complex Cyg OB2 displays similarities to the cluster surrounding R\,136, 
and it would be interesting to compare the cloud temperature and spatial 
distribution of stars in both regions. This suggests that 
the star-forming environment, although producing comparable MF slopes, 
shapes the structural appearance of young star clusters, and thus influences
their long-term survival. While strong mass segregation is observed in 
the high-mass component of each cluster, and especially in the cluster core,
the low-mass population appears constant in shape, decreasing only as expected from
the density profile, over a large range of radii. This may indicate that some 
amount of primordial segregation had to be present in the clusters, even 
if dynamical segregation rapidly condensed the cluster core even more.
Detailed N-body simulations adjusted to the derived cluster characteristiscs
as were performed for the Trapezium cluster by Bonnell \& Davies (1998)
are required to distinguish primordial and dynamical segregation in
NGC\,3603\,YC.

\acknowledgements
We thank our referee Anthony Moffat for his detailed reading of the manuscript 
and his encouraging and valuable comments for both Papers I and II. 
A.\,S. wishes to thank the MPIA and in particular Hans-Walter Rix,
as well as colleagues and friends for the supportive and friendly 
atmosphere during the thesis work, and acknowledges research support 
from the European Southern Observatory via Director General's Discretionary Fund.



\begin{figure*}
\caption{\label{clustersel}
$K_s$ 39 min ISAAC image ($3.\!'4 \times 3.\!'4$, North is up, East left). \newline
The selection of the main cluster center, $R < 33^{''}$ (outer circle), as suggested 
by enhanced stellar density, and of the field population to the East of the cluster
(white boxed area) are shown.
Circles mark radial annuli with $7^{''}$, $20^{''}$, $27^{''}$, and $33^{''}$.}
\end{figure*}


\clearpage

\begin{figure*}
\caption{\label{density}
Density profile of NGC\,3603\,YC. The dotted line indicates the limit
where field star contamination exceeds 10\%.}
\end{figure*}


\begin{figure*}
\caption{\label{n3603incfit}
Recovery fractions from artificial star tests. \newline
The recovery rates are shown for the cluster center and field including
the polynomial fits used in the field star fraction and mass function 
incompleteness correction (left). The radial variation in the completeness
due to varying stellar density is compared for the different annuli (right) 
for which the MF is calculated. Variations are significant only in the innermost
annulus ($7^{''} < R < 20^{''}$). The huge loss of stars in the innermost 
$7^{''}$ reflects the saturation and crowding problems in the cluster core, 
which is excluded from the MF derivation.}
\end{figure*}


\begin{figure*}
\caption{\label{cmd_n3603}
Color-magnitude diagram of the inner $33^{''}$ of NGC\,3603\,YC. \newline
The unsubtracted cluster population is shown in the left panel, the observed
field star population (not yet scaled to the same area)
in the middle, and the subtracted cluster population 
used for mass function derivations in the right panel. 
Field stars were statistically subtracted in color-magnitude bins of 
$0.5\times 0.5$ mag. The 50\% completeness
limit down to which mass functions are fitted is indicated as a dashed line.
Note that the magnitude corresponding to 50\% completeness varies with 
radial distance from the cluster center.}
\end{figure*}


\begin{figure*}
\caption{\label{pmsmssel}
Selection of stars around the average MS (left) and PMS (right) isochrone. \newline
Stars with colors up to 0.25 mag blueward of the PMS and MS isochrones are included,
and redward up to 0.35 mag in the case of MS stars and 0.5 mag in the case of 
PMS stars. The selected stars represent a typical sample used for the MF
derivation (after field subtraction).}
\end{figure*}


\begin{figure*}
\caption{\label{mf_all}
Mass function of NGC\,3603\,YC. \newline
All MFs are derived from the combined MS and PMS population
(Fig.~\ref{pmsmssel}) for an age of 1 Myr, 
distance modulus of 13.9 mag and foreground extinction
of $A_V=4.5$ mag (PMS) and $A_V=4$ mag (MS). Dashed lines are individual
fits for the PMS ($0.4 < M/M_\odot < 4$) and MS ($4 < M/M_\odot < 20$),
and the solid red line is the combined MF fit.
Top panel: MF calculated from simple star counts. 
Middle panel: MF calculated after field subtraction and individual 
dereddening. 
Bottom panel: Mass function derived including binary correction. 
Stars on the binary candidate sequence were shifted down by 
$\Delta J_s=+0.75$ mag, and 30\% of the PMS stars selected randomly 
were treated in the same way.
One equal-mass companion was included for each binary candidate. \newline
The similarity in the combined MF slope is striking despite the 
different procedures used.}
\end{figure*}


\begin{figure*}
\caption{\label{n3603mfrad}
Radial variation in the combined mass function. \newline
Three consecutive annuli with identical area coverage are chosen.
In the combined MF a reasonable fit
could be performed in the two inner annuli up to $16\,M_\odot$, while in the outermost
annulus only the PMS contributes significantly to the stellar population, such that 
the fit proceeds only from 0.4 to $4\,M_\odot$. Despite the observed decrease in the number 
of high-mass stars in the outer annuli, no clear steepening in the MF in the form of a 
steeper slope $\Gamma$ reflects the depletion of high-mass stars at large radii.}
\end{figure*}


\begin{figure*}
\caption{\label{mcum3603}
Cumulative mass distributions of NGC\,3603\,YC. \newline
The CF for the cluster selection $7^{''} < R < 33^{''}$ is shown
in the upper left panel. Theoretical mass distributions are overlaid
(dashed lines) with slopes of $-0.3 > \Gamma > -1.3$ in steps of 0.2, 
the latter corresponding to the Salpeter slope (solid line).
All CFs are normalised to unity at $M = 0.4\,M_\odot$, and the 
absolute number of stars entering each CF is given in the 
upper right corner.
The radial CFs are displayed in the subsequent panels, 
with annuli indicated in the upper right corner. In each radial CF, 
the cluster CF from the first panel is also shown for comparison.
The two outer annuli extending to $65^{''}$ indicate that the 
pre-main sequence population ($M < 4\,M_\odot$) is not truncated 
even at large radii, while the main sequence population is strongly
concentrated towards the cluster center.}
\end{figure*}


\begin{figure*}%
\caption{\label{mcumall3603}
Radial cumulative mass distributions of NGC\,3603\,YC.\newline
This comparison between radial CFs of all annuli shows - in contrast
to the binning-limited MF - a clear trend towards decreasing values
of $\Gamma$ with increasing radius in the inner three annuli.
Beyond $33^{''}$, the increase at the high-mass end is most probably 
due to incomplete field subtraction. On the low-mass end, however, 
the mass distribution correponds very well to the cluster pre-main 
sequence distribution. Indeed, between all 5 annuli, no deviation 
in the distribution can be detected below $3\,M_\odot$ except for 
a slight flattening in the cluster center (top curve). At masses above 
$3\,M_\odot$, a strong decrease in the relative fraction of 
high-mass stars is observed, confirming the mass segregation 
observed in the MFs.}
\end{figure*}


\begin{figure*}
\caption{\label{frac}
Fraction of high- to low-mass stars vs. radius. \newline
The left panel shows the radial change in the number ratio of stars with 
$M \ge 4\,M_\odot$ to stars with $1\,M_\odot < M < 4\,M_\odot$ (left). 
Large symbols represent
the five radial annuli used to study MF variations.
including individual incompleteness corrections for each annulus,
The MS/PMS 
transition is chosen as mass limit between the high- and the low-mass
population, as these two populations are known to be physically
different, such that the ratio simultaneously reveals the 
radial change in the number of MS vs. PMS stars. The strong
decrease in the fraction of high-mass stars reflects the segregation
of massive stars towards the cluster center. The right panel shows
the radial variation in the mass fraction for the same criteria.}
\end{figure*}


\clearpage


\begin{table*}
\caption{\label{binfractab}
Fraction of stars on binary candidate sequence. Only MS stars
are selected. \newline}
\centering
\begin{tabular}{cccc}
\noalign{\bigskip}
$J_s$ mag  & $N_{binary}$ & $N_{single}$ & $N_{binary}/(N_{single}+N_{binary})$ \\
\noalign{\medskip}
\hline
\noalign{\medskip}
12 - 13 mag & 9 & 19 & 0.32 \\
13 - 14 mag & 12 & 38 & 0.24 \\
14 - 14.8 mag & 15 & 42 & 0.27 \\
\noalign{\medskip}
\hline
\noalign{\medskip}
all    & 36 & 99 & 0.28 \\
\end{tabular}
\end{table*}


\begin{table*}
\caption{\label{n3603mftab}
Mass Function derivations for NGC\,3603\,YC. \newline}
\footnotesize
\centering
\begin{tabular}{lccc}
\noalign{\bigskip}
Model number and description & PMS + MS fit & PMS fit & MS fit \\
\noalign{\medskip}
\hline
\noalign{\medskip}
no field subtraction, no dereddening & $\Gamma=-0.91 \pm 0.10$ & $\Gamma=-0.78 \pm 0.32$ & $\Gamma=-0.75 \pm 0.20$ \\
with field subtraction, no deredding & $\Gamma=-0.87 \pm 0.10$ & $\Gamma=-0.72 \pm 0.33$ & $\Gamma=-0.72 \pm 0.25$ \\
with dereddening, no field subtraction & $\Gamma=-0.91 \pm 0.10$ & $\Gamma=-0.85 \pm 0.30$ & $\Gamma=-0.56 \pm 0.20$ \\
with dereddening and field subtraction & $\Gamma=-0.85 \pm 0.10$ & $\Gamma=-0.73 \pm 0.32$ & $\Gamma=-0.57 \pm 0.20$ \\
binary candidates rejected & $\Gamma=-0.97 \pm 0.10$ & $\Gamma=-0.73 \pm 0.33$ & $\Gamma=-0.91 \pm 0.12$ \\
visible binary sequence corrected (MS and transition) & $\Gamma=-1.00 \pm 0.10$ & $\Gamma=-0.73 \pm 0.33$ & $\Gamma=-0.95 \pm 0.22$ \\
visible binary seq. and statistical PMS correction 30\% & $\Gamma=-0.91 \pm 0.08$ & $\Gamma=-0.67 \pm 0.26$ & $\Gamma=-0.84 \pm 0.13$ \\
2 Myr MS, all corrections & $\Gamma=-0.91 \pm 0.08$ & $\Gamma=-0.61 \pm 0.21$ & $\Gamma=-1.03 \pm 0.23$ \\
3 Myr MS, all corrections (fit $< 15\,M_\odot$) & $\Gamma=-0.84 \pm 0.09$ & $\Gamma=-0.64 \pm 0.22$ & $\Gamma=-0.75 \pm 0.10$ \\
\end{tabular}
\end{table*}


\begin{table*}
\caption{\label{n3603mfradtab}
Radial variation of the mass function. \newline} 
\centering
\begin{tabular}{rcl}
\noalign{\bigskip}
\hspace{-1cm} annulus & PMS + MS fit & remark\\
\noalign{\medskip}
\hline
\noalign{\medskip}
 $7^{''} < R < 20^{''}$ & $\Gamma=-0.89 \pm 0.14$ & MS + PMS fit \\
$20^{''} < R < 27^{''}$ & $\Gamma=-1.09 \pm 0.15$ & MS + PMS fit \\
$27^{''} < R < 33^{''}$ & $\Gamma=-0.76 \pm 0.21$ & PMS fit only \\
$33^{''} < R < 46^{''}$ & $\Gamma=-0.99 \pm 0.26$ & PMS fit only \\
$46^{''} < R < 65^{''}$ & $\Gamma=-0.85 \pm 0.23$ & PMS fit only \\
\end{tabular}
\end{table*}


\begin{table*}
\caption{\label{comparetab}
Comparison of young, massive clusters.}
\medskip
\begin{tabular}{lccccccc}
cluster & $M_{total}$ & extent & $r_{core}$ & $\rho_{core}$ & age & MF slope & ref\\
\noalign{\smallskip}
        & $M_\odot$ & pc       &    pc      & $M_\odot\,{\rm pc^{-3}}$ &  Myr & $\Gamma$ & \\
\noalign{\medskip}
\hline
\noalign{\medskip}
Arches & $10^4$ & 1 (?) & 0.2 & $3\cdot 10^5$ & $2 - 3$ & $-0.9 \pm 0.15$ & 3, 4\\
NGC\,3603\,YC & $>\,7\cdot 10^3$ & 4.4 & 0.2 & $10^5$ & $1 - 3$ & $-0.9 \pm 0.1$ & 4 \\
R\,136 & $2\cdot 10^4$ & 4.7 & 0.02 & $5\cdot 10^4$ & $1 - 5$ & $-1.3 \pm 0.1$ & 1\\
Orion & $10^3$ & 3 & 0.2 & $4\cdot 10^4$ & $0.3 - 1$ & $-1.2 \pm 0.1$ & 2\\
\noalign{\medskip}
\hline
\noalign{\medskip}
Antennae & & & & & \\
starbursts & $10^4-10^6$ & $1-10$ & ? & $10^3$ & $1 - 20$ & ? & 5\\
\noalign{\medskip}
Milky Way GCs & $10^4$ - few $10^5$ & few pc & $\approx 1$ & $10^2 - 10^6$ & 10 Gyr & $-1.3$ & various\\ 
\noalign{\medskip}
\hline
\noalign{\medskip}
\end{tabular} \\
{\small 1 - Massey \& Hunter (1998); 2 - Hillenbrand \& Hartmann (1998); 
3 - Figer et al. (1999); 4 - Stolte 2003; 5 - Whitmore et al. (1999).
Average Milky Way globular cluster parameters compiled from various sources.}
\end{table*}

\end{document}